\documentclass{emulateapj}

\usepackage{subfigure,epsfig,longtable,wasysym,amssymb,url}
\begin{document}

\title{Brief Follow-up on Recent Studies of Theia's Accretion}

\author
{Nathan A. Kaib\altaffilmark{1,2} \& Nicolas B. Cowan\altaffilmark{3}}

\altaffiltext{1}{Department of Terrestrial Magnetism, Carnegie Institution for Science, 5241 Broad Branch Road, NW, Washington, DC 20015, USA}
\altaffiltext{2}{HL Dodge Department of Physics \& Astronomy, University of Oklahoma, Norman, OK 73019, USA}
\altaffiltext{3}{Department of Physics \& Astronomy, Amherst College, AC\# 2244, Amherst, MA 01002, USA}

\begin{abstract}

\citet{kaibcowan15} recently used terrestrial planet formation simulations to conclude that the moon-forming impactor (Theia) had only a $\sim$5\% or less chance of having the same oxygen isotope composition as Earth, while \citet{mastro15} used seemingly similar simulations and methods to arrive at a higher value of $\sim$20\% or more. Here we derive the results of both papers from a single set of simulations. Compared to \citet{kaibcowan15}, the analysis of \citet{mastro15} systematically yields more massive Theia analogs and imposes flatter $\Delta^{17}$O gradients across the original protoplanetary disk. Both of these effects diminish isotopic differences between Earth and Theia analogs. While it is notoriously difficult to produce systems resembling our actual terrestrial planets, the analysis of \citet{kaibcowan15} more often selects and analyzes Earth and Mars analogs at orbital locations near the real planets. Given this, we conclude that the greater isotopic differences between Earth and Theia found in \citet{kaibcowan15} better reflect the predictions of terrestrial planet formation models. Finally, although simulation uncertainties and a terrestrial contribution to Moon formation enhance the fraction of Theia analogs consistent with the canonical giant impact hypothesis, this fraction still remains in the 5--8\% range. 

\end{abstract}

\section{Background}

The canonical giant impact formation scenario for Earth's moon predicts that a Mars-mass body, Theia, struck the proto-Earth in a glancing blow which threw material into orbit around Earth and eventually coalesced into the Moon \citep{hartdav75, camward76}. Hydrodynamical modeling of this collision has since shown that the Moon should consist of $\sim$60--90\% of material from Theia, with the rest coming from the proto-Earth \citep{canup04, reufer12}. Yet the Moon's oxygen isotope composition is virtually identical to Earth ($\Delta^{17}$O $\lesssim0.015$\permil) even though the isotopic compositions of nearly all meteorites differ significantly from the Earth \citep{wiechert01, her14}. In particular, Mars has a $\Delta^{17}$O of 0.32\permil. Because simulations of terrestrial planet formation predict that each large rocky body has a wide and stochastic feeding zone, it has been suggested that terrestrial planet formation is unlikely to produce an Earth and Theia that are isotopically similar if other bodies (i.e. Mars) are isotopically distinct \citep{pahlstev07}. 

Terrestrial planet formation simulations are an excellent tool to verify whether we should expect a large compositional difference between Earth and its major impactors. These simulations model the assembly of systems of terrestrial planets via the accretion of $\sim$100--1000 lunar-to-Mars size objects in orbit about the Sun \citep{ray14}. Two recent independent works used suites of these simulations to statistically quantify compositional differences between Earth analog planets and their last major impactors (Theia analogs) \citep{kaibcowan15, mastro15}. While very similar simulations were used, they arrived at significantly different probabilities that Theia's isotopic composition was similar to that of the Earth (and Moon). \citet{kaibcowan15} (hereafter referred to as KC15) concluded that there was a $\lesssim$5\% chance that Theia's feeding zone was similar enough to the Earth's to yield the Moon's observed isotopic composition. On the other hand, \citet{mastro15} (hereafter referred to as MPR15) found a probability of $\gtrsim$20\%.

In this brief follow-up work, we report the cause of this discrepancy. The main differences between the assumptions used in KC15 and MPR15 are:  (1) the criteria employed to select Earth and Mars analog pairs used to calibrate the initial $\Delta^{17}$O distribution in each simulation, and (2) the criteria employed to select Earth and Theia analog pairs used to determine the oxygen isotopic difference between Earth and Theia in each simulation. In KC15, Earth analogs are required to have masses over 0.5 M$_{\oplus}$ and orbits between 0.8 and 1.2 AU, while Mars analogs are assumed to be the next planet outward with a mass greater than 0.05 M$_{\oplus}$. Once the Earth-Mars analog pair is selected, the accretion history of the inner planet (the Earth analog) is searched for an impactor with a mass larger than 0.1 M$_{\oplus}$. If such impactors exist, the last of these is designated a Theia analog, and this body's feeding zone is compared against the Earth analog's. The isotopic composition of the Earth and Theia analogs is then calculated by imposing a $\Delta^{17}$O gradient on the original planetesimals that yields a $\Delta^{17}$O of 0.32\permil~ between the Earth analog and the Mars analog. 

Meanwhile, MPR15 place more emphasis on planet ordering than orbital elements or planet mass to select Earth-Mars analogs. In systems that form 4 or more planets, the 3rd and 4th planets are assumed to represent Earth and Mars, respectively, while in 3-planet systems, Earth and Mars are taken to be the 2nd and 3rd planets out from the Sun. These Earth-Mars pairs are selected regardless of their masses or orbits. These planet pairs are then used to set the primordial $\Delta^{17}$O gradient just as in KC15. Next, any planet in the system is searched for impactors composed of at least 50 particles, and these are taken to be Theia analogs. These impactors need not strike the Earth analog used to set the isotopic gradient and typically strike another planet in the system instead. Finally, the isotopic composition of the Theia analog and its target planet are compared by employing the isotopic gradient imposed by the Earth-Mars analog pair.

\section{Analysis and Results}

To compare the KC15 and MPR15 analyses, we use the 100 CJS and EJS simulations from KC15 since these are virtually equivalent to those employed in MPR15. (It should be noted that the EJS initial conditions are effectively equivalent to the EEJS initial conditions in \citet{ray09}.) KC15 contains a detailed description of the simulations. In KC15, the analysis of CJS and EJS simulations were always performed separately to control for any effect of different giant planet configurations, while MPR15 combined numerous sets of simulations that used a variety of giant planet configurations. Because there was no distinction made between initial conditions sets in MPR15, we combine the CJS and EJS simulations for this follow-up work.

We now apply the analysis described in MPR15 to the combined CJS and EJS simulations to derive the isotopic composition of Theia analogs. The results of this analysis are shown in Figure 1. Here we plot the distribution of $|\Delta ^{17}$O$|$ values for Theia analogs relative to the impacted planet in these simulations. Our modified analysis predicts that 15\% (10 out of 65) of our Theia analogs have $|\Delta^{17}$O$|<0.015$\permil, very near the results of MPR15. Recall that the KC15 results derived from the same simulation sets predict a probability of $\sim$2\% (1 out of 53) that Theia has an Earth-like composition. Thus, we recover the results of both KC15 and MPR15 from the same set of simulations. 

\begin{figure}
\centering
\includegraphics[scale=.44]{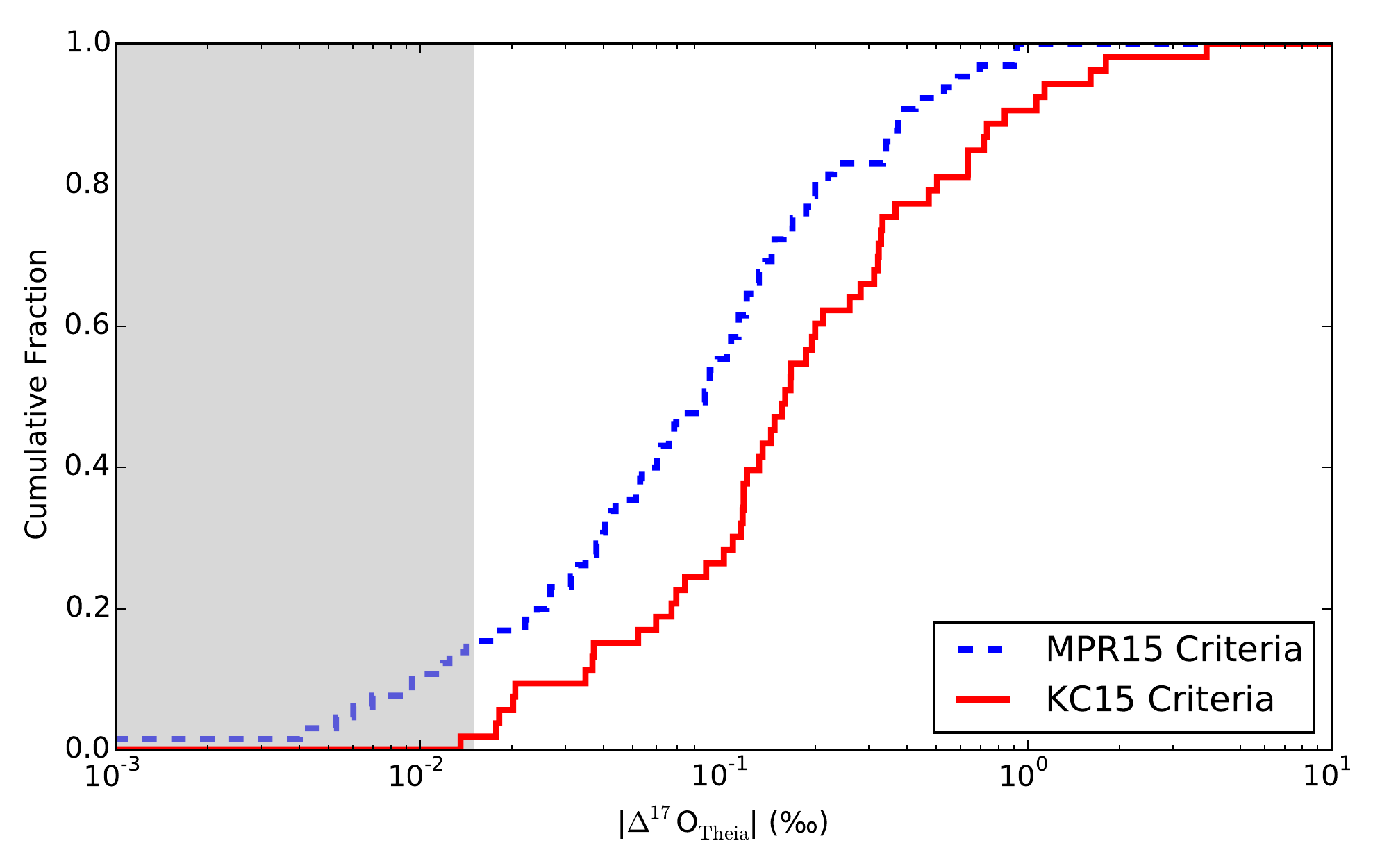}
\caption{The distribution of $|\Delta^{17}$O$|$ for Theia analogs when employing the Earth, Mars, and Theia analog selection criteria of MPR15 ({\it dashed blue}) and when employing the analog selection of KC15 ({\it red solid}). The criteria are applied to the same sets of simulations (CJS and EJS from KC15). The shaded region marks Theia analogs with $|\Delta^{17}$O$|$ values that are less than or equal to that of the Moon (0.015 \permil).}\label{TheiaDists}
\end{figure}

Several different factors cause the differing isotopic distributions for Theia in Figure 1. First, MPR15 require impactors (Theia analogs) to be comprised of at least 50 original particles, whereas KC15 only impose a minimum Theia mass of 0.1 M$_{\oplus}$ and their median Theia analog is comprised of 33 particles. While the 50-particle limit minimizes mass resolution effects, it also causes the impactor masses to be larger in MPR15 than in KC15. In the giant impact scenario, the proto-Earth still accretes nearly all of the impactor, and a more massive impactor pulls the final isotopic composition of the Earth closer to that of Theia. Indeed, when MPR15 relax their 50-particle requirement in their Extended Data Table 2, a clear trend toward more isotopically dissimilar impactors is seen. (In the hit-and-run Moon formation scenario proposed by \citet{reufer12}, the Earth does not actually accrete most of the impactor, and the impactor's influence on the Earth's final composition would be diminished. However, neither MPR15 or KC15 considered this complication.)

When we perform the MPR15 analysis on our simulations, the median mass ratio of Theia to the proto-Earth is 0.35. Meanwhile, the KC15 analysis yields a median mass ratio of 0.25, and only 17\% of KC15 Theia analogs exceed mass ratios of 0.35. Most impacts involving such large mass ratios are typically found for planets well inside 1 AU, nearer the inner disk edge where particle density and accretion rates are higher. Indeed, the planetary targets of the moon-forming collisions selected via the MPR15 criteria have a median semimajor axis of 0.75 AU. This places many target planets near the hard inner disk edge imposed by initial conditions, and these planets may also have inherently different feeding zones compared to the more distant target planets (Earth analogs) required in KC15.

In addition, because the isotopic difference between the Earth and Mars analog pair is always fixed at 0.32\permil, the $\Delta^{17}$O gradient imposed on the initial disk of planetesimals is completely dependent on the separation between the mean formation distance of the Mars analog and the mean formation distance of the Earth analog. A larger separation will produce a flatter gradient, which will in turn yield a smaller difference between the Earth's and Theia's isotopic composition (they are identical if the gradient is completely flat). In many instances, the 4th planet of simulated systems (the Mars analog under the MPR15 criteria) orbits well beyond 2 AU. This often leads to orbital separations of over $\sim$1 AU between the Earth and Mars analogs used to set the $\Delta^{17}$O gradient in MPR15. 

If these widely separated planets also have widely separated mean formation distances, we might expect that some of the success cases of MPR15 come from systems with unusually flat $\Delta^{17}$O gradients. In fact, MPR15 explored the effects of employing steeper $\Delta^{17}$O gradients in their Extended Data Table 2 and found a significant decrease in the fraction of isotopically Earth-like Theia analogs if the disk $\Delta^{17}$O gradients were increased by 25\%. In addition, we find a correlation coefficient of 0.72 when we compare the mean formation separations with the final orbital separations for the MPR15 Earth-Mars pairs. Given this, we may expect that the widely separated Earth-Mars pairs of MPR15 will in fact have widely separated mean formation distances, resulting in flat $\Delta^{17}$O gradients. 

The separations of the mean formation distances for Earth-Mars pairs selected with the KC15 and MPR15 criteria are shown in Figure 2. As expected, we see that the Earth-Mars pairs selected by the MPR15 prescription have a long tail to very high separations not seen among the Earth-Mars pairs selected with the KC15 criteria. As mentioned previously, performing the MPR15 analysis on our simulations yields 10 instances where Theia's $|\Delta^{17}$O$|$ is less than or equal to the Moon's. As Figure 2 shows, 6 of these 10 instances have Earth-Mars pairs that lie in the tail of very high Earth-Mars separations, which require flat $\Delta^{17}$O gradients. This suggests that the different distributions of Earth-Mars analog formation distances is a primary reason for the increased likelihood that Theia has an Earth-like isotopic composition in MPR15. 

\begin{figure}
\centering
\includegraphics[scale=.44]{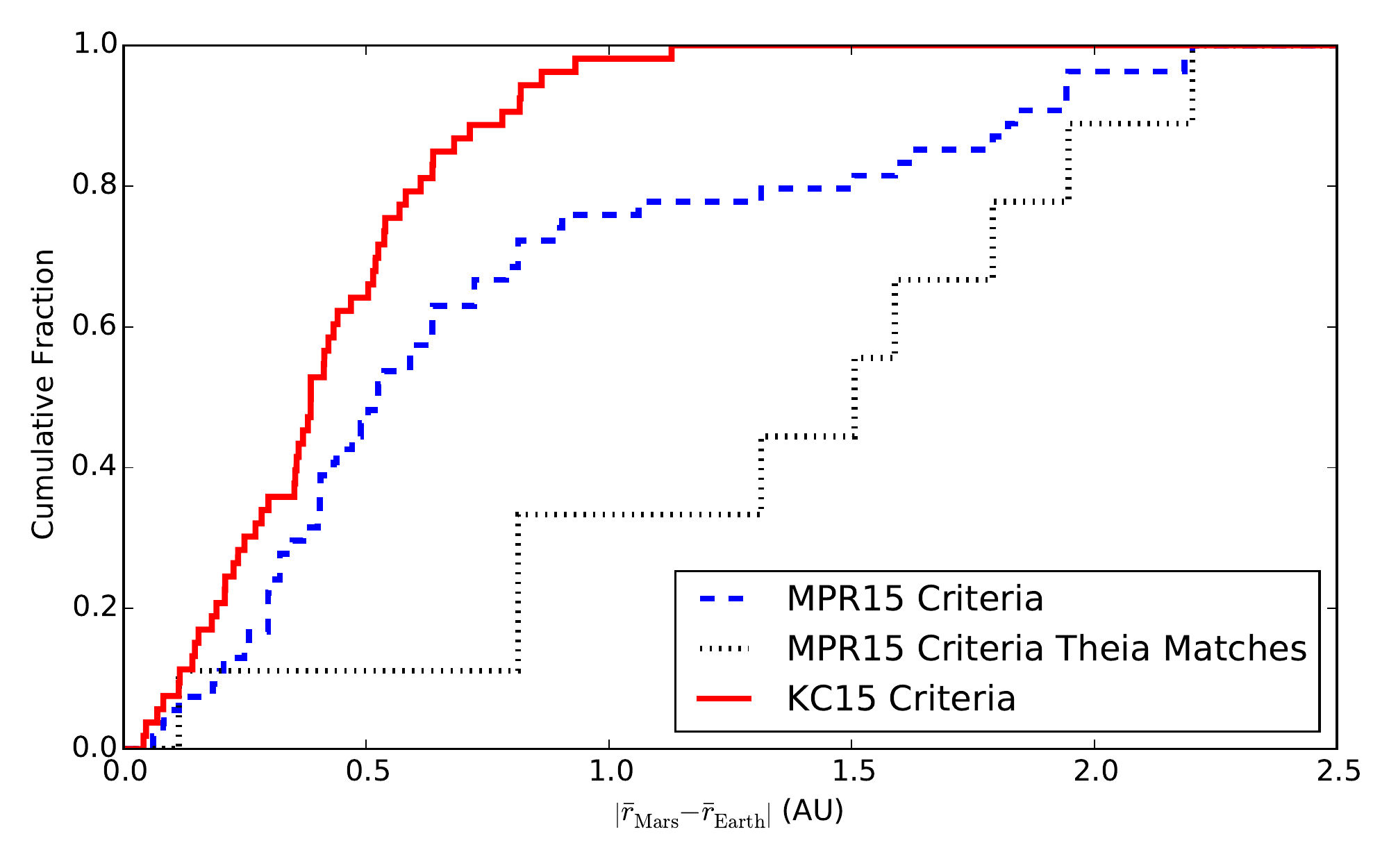}
\caption{The distribution of separations between the mean formation radii of Mars analogs and Earth analogs as selected by the criteria of MPR15 ({\it dashed blue}) and by the criteria of KC15 ({\it red solid}). The distribution of Earth-Mars separations is also shown for simulations that yielded Theia analogs with $|\Delta^{17}$O$|$ values equal to or less than that of the Moon when employing the MPR15 criteria ({\it dotted black}). }\label{a sep}
\end{figure}

The large separations between the mean formation distances of Earth analogs and Mars analogs seem to only occur in systems where at least one of the two analog planets has a very small mass. In Figure 3 we plot the formation distance separation against the minimum planet mass of each Earth-Mars analog pair. Here we see that the separation between Earth's and Mars' formation distance is only above 1 AU if at least one of the analog planets has a mass well below 0.1 M$_{\oplus}$. These very low mass planets are comprised of one to a few particles, and it has been shown that individual particles can migrate very large distances during terrestrial planet formation simulations \citep{cham01, ray04, morish10}. This causes the large formation distance separations and the flat $\Delta^{17}$O gradients that are consequently imposed. If we only consider cases where both Earth and Mars analogs have masses above 0.06 M$_{\oplus}$ (approximately the smallest planetary mass seen in our Solar System) then the fraction of MPR15 Theia analogs with $|\Delta^{17}$O$| < 0.015$\permil~ drops to just 3\% (1 out of 31 cases), nearly the same as the KC15 results.

It should be noted, however, that Mars analogs should in fact be substantially smaller than the Earth analogs, and no such requirement was made in KC15. If we use the KC15 Earth-Mars pairs and now require that the Earth analog is at least twice as massive as the Mars analog, our usable sample of simulations drops from 53 to 18. Of these 18 systems, we still only have one Theia analog with $|\Delta^{17}$O$|<0.015$\permil~ (a 5.5\% success rate), and the median value of $|\Delta^{17}$O$|$ actually increases slightly (0.16\permil~ vs. 0.18\permil). Thus, the results of KC15 do not seem to be due to having selected Mars analogs that are unrealistically massive.

\begin{figure}
\centering
\includegraphics[scale=.44]{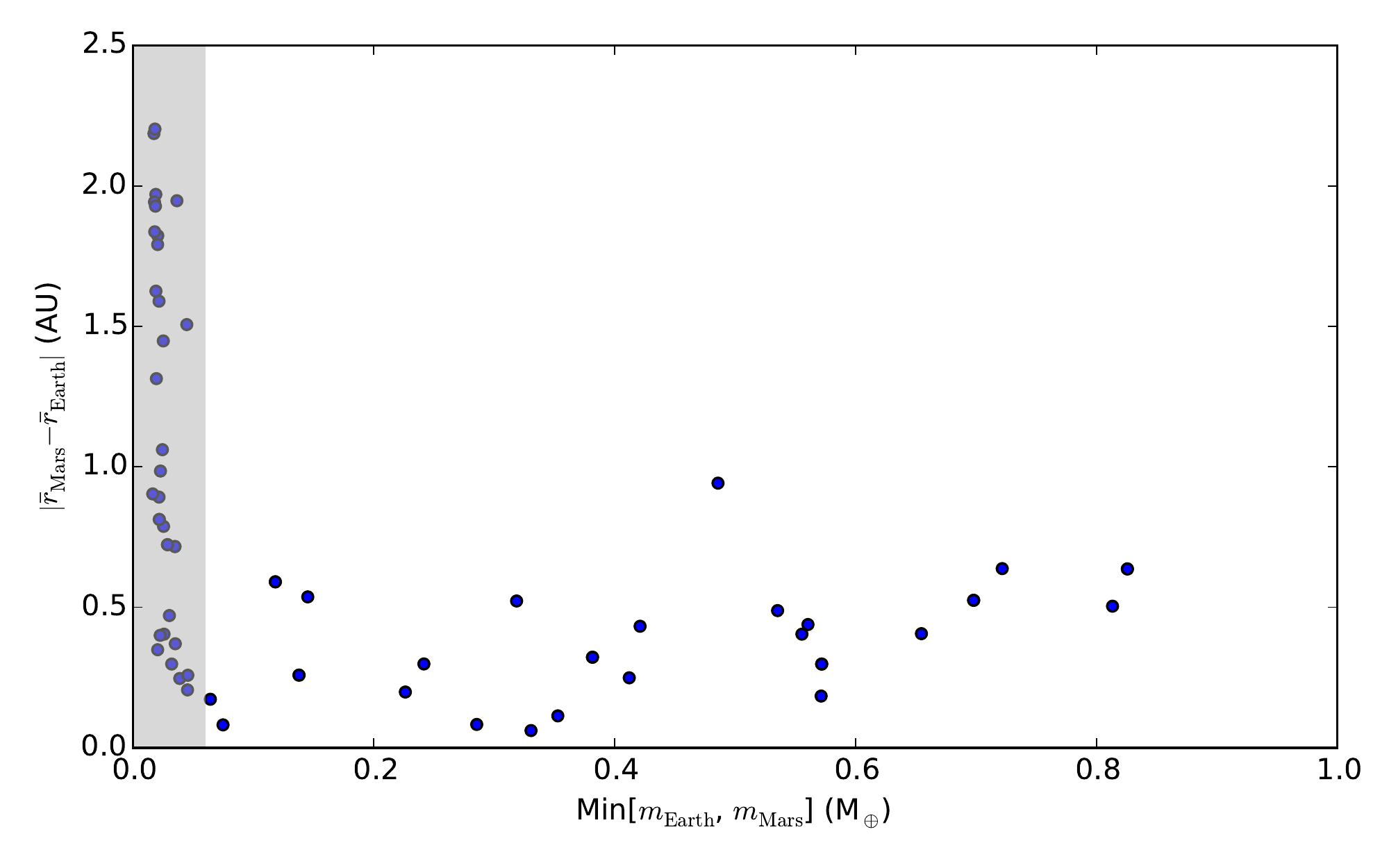}
\caption{The separation between the mean formation distances of Mars analogs and Earth analogs as selected by the criteria of MPR15 ({\it dashed blue}) are plotted against the minimum planet mass of each Earth-Mars pair. The shaded region highlights systems where the mass of Earth and/or Mars is below 0.06 M$_{\oplus}$. }\label{a sep}
\end{figure}

\subsection{Simulation Uncertainties}

The stochastic nature of planet formation is well captured by numerical experiments: if a simulation is repeated with virtually identical initial conditions, planets form at different locations and have different feeding zones.  This introduces an uncertainty in the results of any individual simulation, and is precisely the reason that KC15 and MPR15 ran large ensembles of simulations to quantify the likelihood of the Moon-forming impact.  

The ``granular'' nature of N-body simulations also introduces an uncertainty in the planetary composition. Our simulations contain 1000 planetesimals to represent small bodies, but the real number may have been orders of magnitude higher during Solar System formation. The feeding zone of each simulated protoplanet is likely constructed from a much smaller number of planetesimals than in the actual solar system. 

Granularity is a consequence of computational limitations and can artificially skew our results. Even if the true feeding zones of a Theia analog and an Earth analog were identical, our simulations could result in a non-zero $|\Delta^{17}$O$|$ due to the small number of simulated planetesimals. Alternatively, the coarse mass resolution of our simulations could produce two planets with nearly identical compositions where a higher-resolution simulation would have shown that they have significantly different feeding zones. 



Disentangling granularity error from natural stochasticity is difficult. 
MPR15 estimate the uncertainty in each simulated major body's mean formation distance using the standard error of the mean (SEM). The SEM values for Earth and Theia analogs are then added in quadrature and translated to a $\Delta^{17}$O uncertainty.  When simulated $|\Delta^{17}$O$|$ values are below 0.015\permil + the SEM uncertainty, MPR15 deem that the simulation successfully explained the Moon's isotopic composition.  This only makes sense, however, if one presumes that the feeding zones of Earth and Theia are inherently identical and that simulation granularity artificially increased compositional differences.  


We instead estimate the granularity error of our simulations by resampling each major body's planetesimal distribution via bootstrap. If a simulated planet accreted $N$ planetesimals from different semi-major axes, then we draw $N$ times from that planetesimal distribution, with the possibility of repeated selections. This produces a new distribution of $N$ planetesimals.  By repeating this process 1000 times for each body, we generate a distribution of feeding zone locations for each simulated body.  Bodies made of a few planetesimals from disparate semi-major axes will have higher variance than bodies built from many nearby planetesimals. The standard deviation of the feeding zone location for each body is an estimate of the granularity uncertainty, and is typically within a few percent of the SEM value. 

 
We only perform bootstrapping and SEM for distributions of planetesimals, which are all equal-mass particles. The number of simulated embryos is comparable to expected number in the real solar system, so our embryo population should not suffer from the granularity issues. It is not clear whether or not MPR15 make a distinction between embryos and planetesimals in their SEM calculations.

We use bootstrap resampling to predict---for each simulation---the probability that the feeding zones yield a Theia analog with $|\Delta^{17}$O$|\le0.015$\permil.  For each individual set of Earth-Mars-Theia feeding zones, we use bootstrap resampling to generate a distribution of 1000 $\Delta^{17}$O gradients and $\Delta^{17}$O values for each individual simulation. We then sum up the total probability that $|\Delta^{17}$O$|<0.015$\permil~ for our ensemble of simulated Theia analogs. Unlike the raw simulation results, these results will account for the coarse mass resolution of our simulations. This analysis predicts a $\sim$4.8\% probability that Theia will have $|\Delta^{17}$O$|<0.015$\permil.  This number is 2.5$\times$ larger than the predictions of our raw simulation results, but still significantly smaller than the prediction of MPR15.

Finally, it is worth noting that both SEM and bootstrap resampling implicitly assume that each accreted object represents an independent sample of a protoplanet's ``true'' underlying feeding zone. During accretion, however, small bodies often merge with each other before being accreted onto even larger bodies. Since much of a protoplanet's feeding zone is derived from conglomerations of associated bodies, the initial locations of accreted planetesimals are likely correlated. It is therefore possible that neither SEM nor bootstrapping properly quantifies the uncertainty arising from using too few planetesimals.

\subsection{Earth's Contribution to the Moon}

In KC15, it was assumed that lunar material is almost entirely derived from Theia instead of the Earth. In reality, Earth always contributes some fraction of material toward the Moon's assembly. In the canonical giant impact hypothesis, the Earth fraction is typically around 15--25\% \citep{canup04}. In alternative scenarios, such as hit-and-run collisions, high velocity impacts, or very massive impactors, the terrestrial fraction of the Moon's material can exceed 40\% \citep{reufer12, cukstew12, canup12}. Considering a significant terrestrial contribution to the Moon's mass allows Theia analogs with larger $|\Delta^{17}$O$|$ values to explain the Moon's isotopic composition since the $|\Delta^{17}$O$|$ will be scaled down by the compositional fraction that Theia provides (1 minus the terrestrial fraction) when Theia material is incorporated into the Moon. In Figure 4, we examine how the fraction of Theia analogs that successfully explain the lunar isotopic composition varies as we consider greater contributions from the Earth. Using our $|\Delta^{17}$O$|$ values generated from bootstrap resampling, we find that our Theia success rate moves from 4.8\% to 8.4\% as we consider Earth contributions up to 40\% of the Moon's mass. When looking at the raw simulation data instead of our resamplings, one finds an increase from $\sim$2\% with no terrestrial contribution up to $\sim$9.5\% if the Earth supplies 40\% of the Moon's material.

Unlike KC15, MPR15 did consider scenarios in which the Moon had a significant terrestrial component and found that a terrestrial contribution of 40\% could increase the success rate of Theia analogs up to 25\% or 55\%, depending on whether their SEM-derived uncertainties were being considered. As shown in Figure 4C, KC15 finds success rates that are significantly lower than either of these values. We attribute this difference to the Earth, Mars, and Theia analog selection criteria employed in KC15 vs MPR15.

\begin{figure}
\centering
\includegraphics[scale=.44]{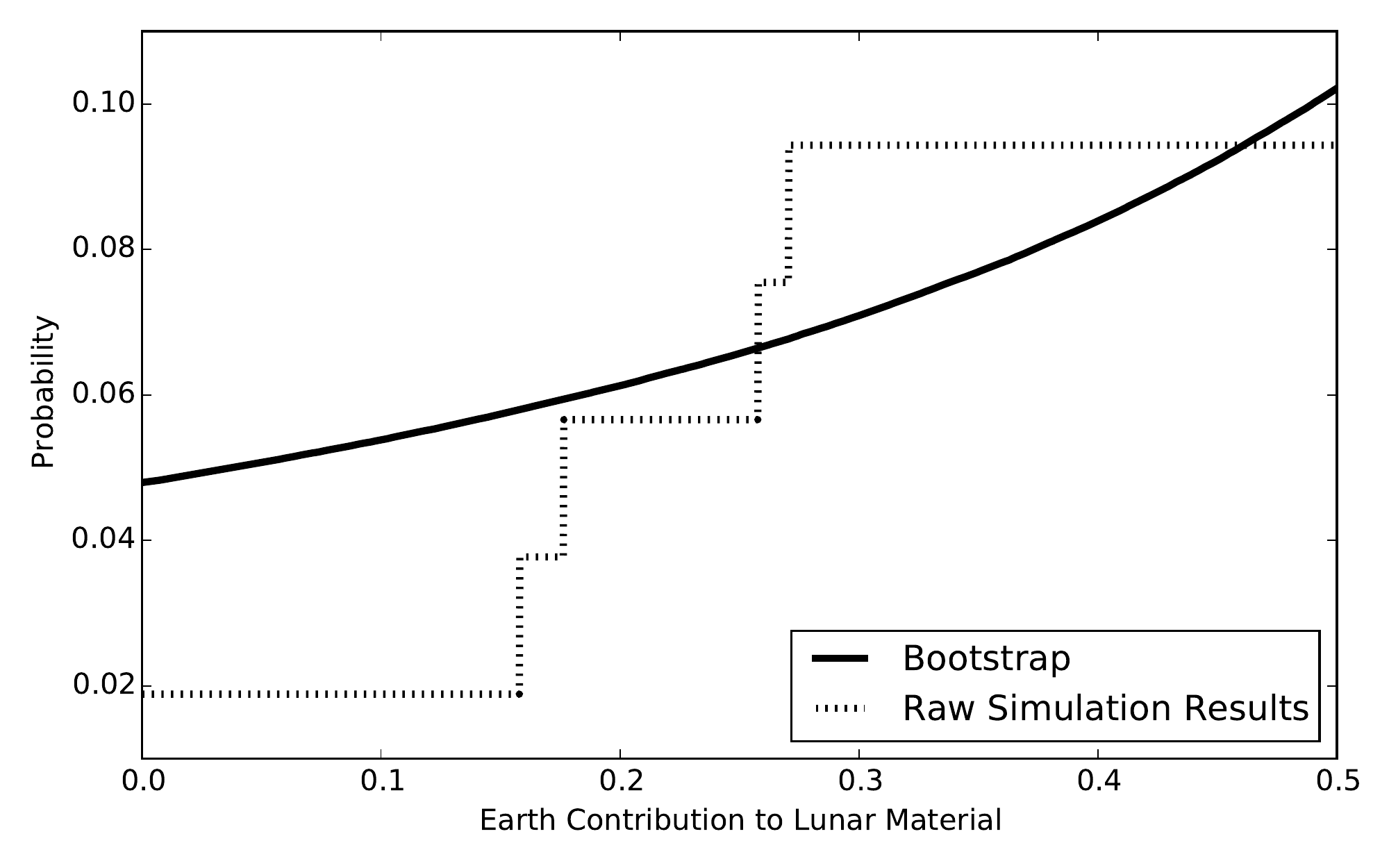}
\caption{The probability that Theia will have a $\Delta^{17}$O value capable of producing a moon with $|\Delta^{17}$O$|<0.015$\permil~ is plotted against the assumed fraction of lunar material derived from the Earth. This probability is calculated from our bootstrapped $\Delta^{17}$O ({\it solid black}) distribution, which accounts for granularity uncertainty, and from our raw simulation results ({\it dotted black}).}\label{fit}
\end{figure}

\section{Discussion and Conclusions}

Although recent terrestrial planet formation models have shown more promise in producing Martian analogs \citep{walsh11, izi14}, many of the simulations employed in KC15 and MPR15 have difficulty reproducing the low mass of Mars \citep{ray09}. Because of this, it is not obvious how to designate Earth and Mars analogs. However, we believe the KC15 criteria are more self-consistent for several reasons. First, the target of the moon-forming collision is always required to be the same planet used in calibrating the initial $\Delta^{17}$O of simulation particles, whereas the target planets selected under the criteria of MPR15 often orbit inside Venus' orbit. In addition, to minimize simulation resolution effects, MPR15 require Theia analogs to be accreted from at least 50 particles, while the median Theia analog in KC15 is comprised of 33 particles. However, this 50-particle limit forces Theia analogs to always have larger masses than that predicted in the canonical giant impact hypothesis. Finally, although the masses of Mars analogs are often too large in KC15, all of the simulations that are used in this work have a planet orbiting near 1.5 AU, just as the Solar System does, and this is used to calibrate the initial $\Delta^{17}$O distribution. Meanwhile, the Earth-Mars pairs selected with the MPR15 criteria can often orbit in the 2-3 AU region and have masses similar to or less than Mercury, something unseen in our Solar System. Flawed as these simulations may be at replicating the detailed structure of the inner Solar System, they do predict that the last major impactor on a massive planet near 1 AU is quite unlikely to have the same isotopic composition as the planet itself, when there is also a planetary mass body near 1.5 AU with a Mars-like composition. 

MPR15 correctly point out that the feeding zones derived from terrestrial planet formation simulations are artificially influenced by the mass resolution of simulations. This increases the uncertainty in feeding zone locations, and hence $\Delta^{17}$O values. When we account for this uncertainty, the fraction of Theia analogs capable of explaining the Moon's isotopic similarity to the Earth increases from 2\% to 4.8\%. Accounting for the possibility that the Moon could contain up to 40\% terrestrial material further increases the successful fraction of Theia analogs to 8.4\%. However, this is still well below the $\sim$50\% success rates reported by MPR15. Given this, we conclude that standard models of the final assembly of terrestrial planets predict there is a $\sim$5\% chance that the canonical giant impact hypothesis can account for the isotopic composition of the Moon.

\section{Acknowledgements}

We thank John Chambers, Alessandra Mastrobuono-Battisti, and Hagai Perets for useful discussions. We also thank Sean Raymond and an anonymous referee for helpful criticisms and suggestions that improved the quality of this work.

\bibliographystyle{model1-num-names}
\bibliography{MoonComments}

\end{document}